# Unlabelled Far-field Deeply Subwavelength Superoscillatory Imaging (DSSI)

T. Pu[1$], V. Savinov[1], G. Yuan[2], N. Papasimakis[1], N. I. Zheludev[1,2*]

[1]Optoelectronics Research Centre and Centre for Photonic Metamaterials, University of Southampton, Southampton SO17 1BJ, United Kingdom

[2]Centre for Disruptive Photonic Technologies, The Photonics Institute, School of Physical and Mathematical Sciences, Nanyang Technological University, 637371 Singapore

*Correspondence to: zheludev@soton.ac.uk

[$]On leave from: Key Laboratory of Microelectronics Devices & Integrated Technology, Institute of Microelectronics, Chinese Academy of Sciences, Beijing 100029, China.

**Abstract:** Recently it was reported that deeply subwavelength features of free space superoscillatory electromagnetic fields can be observed experimentally and used in optical metrology with nanoscale resolution (*1*). Here we introduce a new type of imaging, termed Deeply Subwavelength Superoscillatory Imaging (DSSI), that reveals the fine structure of a physical object through its far-field scattering pattern under superoscillatory illumination. The object is reconstructed from intensity profiles of scattered light recorded for different positions of the object in the superoscillatory field. The reconstruction is performed with a convolutional neural network trained on a large number of scattering events. We show that DSSI offers resolution far beyond the conventional "diffraction limit" even if low dynamic range optical detection is used. In modelling experiments, a dimer comprising two subwavelength opaque particles is imaged with a resolution exceeding $\lambda/200$. We show that such resolution levels are tolerant to noise and achievable with low-dynamic range photo-detectors.

The development of label-free far-field super-resolution imaging, beyond the half-wavelength limit of the conventional microscope, remains one of the main challenges for science and technology. Indeed, the ability to image at the nanometer scale using visible light will open unprecedented opportunities in the study of biochemical, biomedical, and material sciences, as well as nanotechnology. However, despite persistent research efforts, deep subwavelength resolution is only possible using techniques, such as STED (*2*) and SMLM/STORM (*3, 4*) that require labelling of samples with luminescent material that is not acceptable for many biomedical and nanotechnology applications (e.g. imaging of semiconductor chips). Here we introduce a non-invasive optical imaging technique with resolution exceeding $\lambda/100$. Our method allows far-field, non-contact, label-free optical imaging with close to molecular resolution by exploiting advances in singularity optics and machine learning. We term our method Deeply Subwavelength Superoscillatory Imaging (DSSI). The extreme resolution of our technique comes from the analysis of a large amount of information about the object that is gathered by registering the results of multiple interactions with highly structured light beams at different illumination conditions.

Recent interferometric experiments (*5*) confirmed the long existing theoretical observations (*6*) that complex coherent optical beams could contain highly localized intensity hotspots and zones of energy backflow. They also revealed that phase in such optical fields could change on a distance orders of magnitude smaller than the wavelength of light. Such complex optical fields, known as superoscillatory fields, can be generated through interference of multiple waves diffracted on a grating (*7*) or purposely designed masks (*8*). Here we show that DSSI can reveal the fine subwavelength structure of an object through recording intensity profiles of a large number of far-field scattering patterns taken under superoscillatory illumination. Our approach exploits the rapid spatial variations of the illuminating optical field as even small displacements of a deeply subwavelength object placed near a phase singularity can significantly change the far-field pattern of light scattered by the object (see Fig. S1 in Supplementary Information).

In contrast to the conventional microscope that forms the image of the object in a single exposure and is limited in resolution at about half wavelength of the light used for illuminating the object, the DSSI technique requires post-processing of multiple scattering field patterns for imaging. We show that a convolutional neural network trained on a large number of scattering events prior to the imaging act is a very powerful tool that can reliably retrieve information about the object with deeply subwavelength resolution. Moreover, our analysis shows that high resolution is tolerant to noise and achievable with low-dynamic range photo-detectors.

A direct reconstruction of an imaging target is possible only if the intensity and phase of the scattered field is known on a closed surface encompassing the object with infinite precision (Kirchhoff–Helmholtz integral). Recently developed monolithic optical micro-interferometry (*5*) can, in principle, detect the phase of the field everywhere around an isolated scattering object, but the technique is extremely challenging for routine microscopy (*9*). Although the field intensity of scattered light is much easier to measure, as no interferometry is required, image reconstruction from only the intensity profiles is an ill-posed inverse problem (*9*). Different iterative feedback algorithms have been developed enabling the reconstruction of an image from intensity of scattering patterns of optical, deep UV and X-ray radiation with resolution essentially limited by the wavelength of the illuminating light in most cases (*10-15*), and around 5-times higher when compressed sensing techniques for imaging sparse objects are used (*16-18*). Also, images taken by machine vision cameras have been enhanced by artificial intelligence algorithms but without demonstrating subwavelength resolution(*19, 20*).

Here we show that object reconstruction with deep subwavelength resolution and high finesse is achievable by detecting only the intensity profile of the scattered light. We also show that higher resolution is achievable if instead of conventional plane wave illumination a superoscillatory illumination is used. We reconstruct the key spatial parameters of the object from the intensity profiles of scattered light with a deep learning neural network trained on a large number of scattering events. We illustrate the Deeply Subwavelength Superoscillatory Imaging (DSSI) by modelling the imaging of a dimer, a pair of randomly positioned subwavelength particles. Imaging a dimer is a simple, but important task that appears often in bio-imaging and nanotechnology (e.g. imaging of cell division or nanoantennas).

In our modelling of proof-of-principle one-dimensional imaging experiments (see Fig. 1), we image a dimer consisting of two totally absorbing, non-scattering elements of widths A and C with gap B between them. The dimer's location from the center of the object plane is represented by distance D. The scattered light is detected by an intensity detector array that is placed at a distance of $10\lambda$ from the object plane, over its center. Here $\lambda$ is the wavelength of the free-space radiation used in the modelling. We assume that the detector array is $10\lambda$ long. Since the scattered field reaching the detector array is formed by free-space propagating waves, in a real experiment it can be imaged at any magnification without loss of resolution, by simply adjusting the magnification to ensure that the detector pixels smaller than the required resolution, as has already been demonstrated in a real experiment (*5*). We therefore assume that the array can image the intensity profile of the diffracted/scattered light without any limitations to spatial resolution and conduct our modelling for a detector array containing five thousand pixels.

We consider two closely related situations, where the position of the dimer at the imaging plane is either known or unknown. We assume that the dimer with unknown position is located anywhere within a chosen interval. In the former case, the imaging process returns the dimensions of A, B, C of the dimer, whereas in the latter the position D is returned in addition to the dimer dimensions. We used illumination with a superoscillatory wavefront generated by a planar Pancharatnam-Berry phase metasurface that was developed in the experimental work reported in Ref. (*1*). It creates a superoscillatory subwavelength hotspot at a distance $z=12.5\lambda$ from the plane of the metasurface. The hotspots are flanked by zones of high phase gradient (see Fig. 1). For comparison, we also used plane wave illumination.

In the imaging process, the superoscillatory field generator is scanned across the object by steadily moving the superoscillatory hotspot, at intervals of $\lambda/5$, from the $-\lambda$ position to the $+\lambda$ position in the object plane. For each position of the hotspot the detector records a diffraction pattern. The full set of diffraction patterns is analyzed by a Convolutional Neural Network (*21*) to retrieve information about the object. The network contains three convolution layers with 32-5×5, 32-3×3, 64-3×3 and 32-1×1 kernels, correspondingly, and three fully connected layers with 128, 32, 4 neurons, respectively. The first three convolution layers are separately followed by a pooling layer with 1×4, 1×8, 1×4 kernels with Rectified Linear Unit activation function. The network was trained with the Adam stochastic optimization method (*22*) and mean absolute error loss function, aimed at improving the retrieval of the dimer geometrical parameters, i.e. constants A, B, C and D. The training data set contained 20,000 samples. It was generated by creating dimers of random sizes and placing them on the object plane with the dimer center coordinate D randomly chosen in the interval from $-\lambda/2$ to $\lambda/2$ (in the case of unknown dimer position). The widths of the dimer components (A and C) and the gap between them (B) were independently and randomly chosen between $0.002\lambda$ and $\lambda$. The diffraction pattern on the detector array was then calculated by the Fourier propagation method (*23*) for the transverse component of the electric field.

The results of our imaging experiments are presented in Fig. 2. They demonstrate that the dimensions of the dimer and its position can be retrieved accurately. Indeed, on Fig. 2 the solid red and blue lines correspond to the median of the true values as a function of the retrieved value, whereas the black solid line is the bisector of the first quadrant ($y=x$) representing perfect agreement between true and retrieved values. A departure of the median from the bisector represents a systematic bias in the retrieval process. When the position of the dimer is known, we obtain remarkably accurate retrieval of all dimensions both for plane wave (red lines) and superoscillatory illumination (blue lines), with the systematic bias of $\sim\lambda/100$ or smaller. Here, superoscillatory illumination gives similar results to plane wave illumination for the size of dimer's element A, but provides over a factor of x2 smaller systematic bias for dimer gap B. When the dimer position is *a priori* unknown, the systematic bias increases but remains sub-$\lambda/100$ with superoscillatory illumination still giving better results for retrieval of dimer gap B and position D than plane wave illumination. See Supplementary Information for more details.

A parameter that is of most interest in real experiments is the spread of true values for a given measurement result, which defines the resolving power of the method. Here, we quantify the resolving power of DSSI by calculating the interquartile range (IQR) of the distribution of true values given a retrieved value (see Fig. 3). We found that the IQR does not vary significantly with the dimensions of the dimer and use its mean value as resolution. Remarkably, in the case of known position and superoscillatory illumination, the resolution of the imaging process exceeds $\lambda/200$ for all dimer parameters. When the position of the dimer is not known, the resolution decreases to $\sim\lambda/80$ for superoscillatory illumination. In both cases, superoscillatory illumination provides a resolution enhancement of >50% over plane wave illumination (see Supplementary Information). The imaging results presented on Fig. 3 were obtained by using the intensity of the diffracted pattern resolved with 16-bit precision corresponding to a dynamic range of 96dB. Here the dynamic range is defined as $10 \cdot \log_{10}(I_{max}/I_{min})$, where $I_{min}$ and $I_{max}$ are the minimum and maximum intensity levels that can be recorded. Although such dynamic range is achievable with high-quality photodetectors, the resolution of the method is weakly dependent on the dynamic range of the detectors. To illustrate this, the detector's dynamic range was deliberately reduced by rounding readings to lower values (Fig. 3f). Nevertheless, resolution at the $\lambda/100$ scale is achieved even for 40 dB dynamic range, whereas typical photodiode values are well above the 60 dB level.

The extraordinary deeply subwavelength level of resolution reported here, at the level better than $\lambda/100$, exceeds the Abbe "diffraction limit" of resolution ($\sim\lambda/2$) by two orders of magnitude. We argue that several factors contribute to this improvement:

1. Recording of multiple scattering patterns provides much more information on the imaged object for the retrieval process than what is available in the lens-generated single image for which the Abbe limit has been derived;
2. The deep learning process involving a neural network trained on a large data set creates a powerful and accurate deconvolution mechanism without using explicit information on the phase of the detected signals;
3. Sparsity of the object and prior knowledge about the object (dimer of unknown size and location) help the retrieval process, similarly to how sparsity helps 'blind' compressed sensing techniques (*17*);
4. Superoscillatory illumination ensures much higher sensitivity of the pattern of scattered light to small features of the imaged object than conventional illumination.

The last argument requires a more detailed comment. It shall be noted that superoscillatory fields contain zones of rapid phase gradient and high local wave vectors leading to high spatial resolution through Fourier connection between spatial and reciprocal space. Although this fact is reassuring, its full implication is difficult to analyze in the context of the multiple exposures and the neural network deconvolution used in the DSSI technique. Instead, on Fig. 4 we illustrate the sensitivity of the scattered field pattern on placing a small absorbing nanoparticle in the illuminating field. The nanoparticle, only $\lambda/1000$ in size, is positioned on the object plane at coordinate $x_0$ (see Fig. 1 and Figs. 4e-f) and illuminated with coherent light of wavelength $\lambda$. The intensity of the scattered light is detected at a distance $z = 10\lambda$ from the nanoparticle at points with coordinates ($x, z = 10\lambda$). Maps (a) and (c) illustrate sensitivity of scattering to the presence of the particle in the illuminating field for plane wave and superoscillatory illumination, respectively. They show the normalized change of the intensity of scattered light (colormap, logarithmic scale) as a function of the particle position $x_0$ on the object plane and the detector's coordinate $x$. Maps (b) and (d) illustrate sensitivity of scattering to small displacements of the particle. They show the normalized change of the scattered field intensity (colormap, logarithmic scale) on displacing the particle with step of $\lambda/2000$ along the object plane with the particle initially located at $x_0$. From Fig. 4 it follows that scattering of the superoscillatory field is two to three orders more sensitive than in the case of plane wave illumination to the presence and repositioning of the nanoparticle, which we attribute to the presence of high intensity and phase gradients in the superoscillatory field. In particular (see Fig 4c), placing the particle anywhere apart from the very narrow subwavelength singularity zone (black horizontally extended area indicated by green dotted line) results in strong change of intensity across the detector plane. Fig. 4d shows that when the nanoparticle is repositioned away from the singularity point in the object plane, a very narrow, deeply subwavelength zone is created on the detector plane where no change of intensity is taking place. These features can be used to accurately retrieve the particle position.

The DSSI reported here shall be compared with the superoscillatory imaging technique that uses a subwavelength intensity hotspot for illumination of the object. The image is reconstructed point-by-point by scanning the hotspot against the object while the scattered light is detected with a lens through a confocal aperture (*24*). In the superoscillatory imaging technique, the size of the superoscillatory hotspot determines the resolution of the technique (*25*). Although, in principle, the superoscillatory hotspot can be arbitrary small, intensity in the hotspot rapidly drops with its size, and resolution better than $\lambda/6$ has never been experimentally demonstrated. As we have shown here, the resolution of DSSI is orders of magnitude better than imaging based on superoscillatory illumination with confocal detection, which is explainable by the availability of additional information for the imaging process.

In practical terms, the main challenge in experimental implementation of Deeply Subwavelength Superoscillatory Imaging would be in creating reliable and trustworthy training sets for the neural network. Such training sets can be generated by computer modelling and great care shall be needed to accurately match characteristics of the imaging apparatus to the model. Alternatively, the training set can be generated *in situ* of the microscope, which will require a large physical library of random objects to be generated and imaged. Resolution will also be constrained by the signal to noise ratio at the detector. However, our results (see Fig. S3 in Supplementary Information) indicate a remarkable resilience of the method, where even in the case of 5% noise, a dimer can be imaged at a resolution of $\sim\lambda/71$ for the element size, $\sim\lambda/77$ for the gap, and $\sim\lambda/92$ for the position. Finally, an additional restriction is imposed by stability and the precision of optomechanical components, which will have to be in the Angstrom scale (as is the case for atomic force microscopy).

In conclusion, we have introduced the new concept of Deeply Subwavelength Superoscillatory Imaging (DSSI), which employs artificial intelligence to retrieve, with resolution exceeding

λ/200, parameters of a physical object from its scattering pattern upon superoscillatory illumination. Although so far the concept has been demonstrated for one-dimensional imaging, it can be extended to two- and three-dimensional objects, as well as objects of *a priori* unknown shape. The technique does not require labelling of the sample with luminescent materials nor intense laser illumination and is resilient to noise. The technique promises far-reaching consequences across a number of disciplines, such as biomedical sciences, materials science and nanotechnology.

**Acknowledgments.** This work was supported by the Singapore Ministry of Education (Grant No. MOE2016-T3-1-006), the Agency for Science, Technology and Research (A*STAR) Singapore (Grant No. SERC A1685b0005), the Engineering and Physical Sciences Research Council UK (Grants No. EP/N00762X/1 and No. EP/M0091221), and the European Research Council (Advanced grant FLEET-786851). T.P. acknowledges the support of the Chinese Scholarship Council (CSC No. 201804910540).

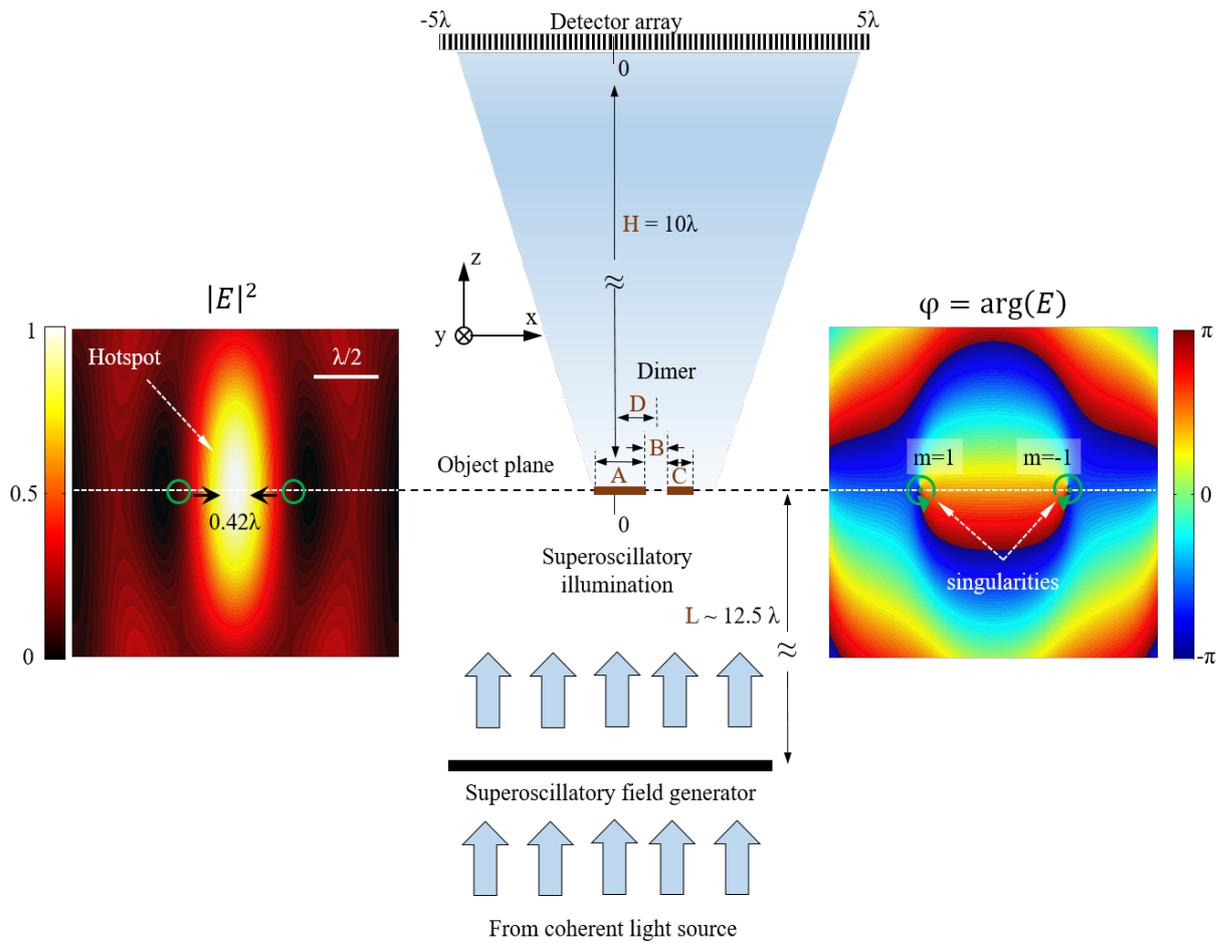

**Fig. 1. Deeply Subwavelength Superoscillatory Imaging (DSSI) schematics.** The imaged object (a dimer A-B-C) is illuminated with a superoscillatory light field. The intensity profile of the diffraction pattern resulting from scattering of the superoscillatory light field on the imaged object is detected by the detector array. A number of different diffraction patterns are recorded when the illuminating field is scanned against the object. Left and right panels show maps of intensity and phase profiles of the illuminating field and indicate the presence of hotspots and phase singularities, where *m* indicates the winding number of the singularity.

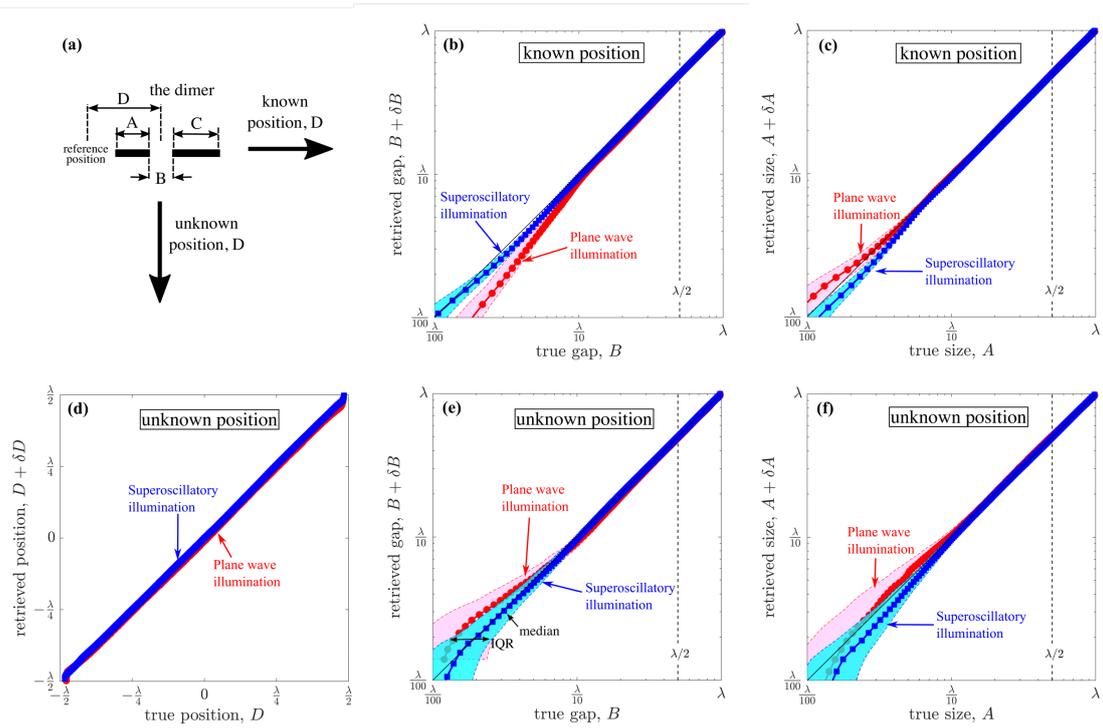

**Fig. 2. Deeply Subwavelength Superoscillatory Imaging of a dimer.** The dimer consists of two elements with different sizes A and C separated by a gap (edge-to-edge) B (panel (a)). It is positioned in the object plane at distance D from the $x=0$ points of the object plane (see Fig. 1). Two different regimes are presented, where the dimer position is either **known** (fixed at D=0) (panels (b,c)), or **unknown** (panels (d-f)). Panels (b, c) show the retrieved values of A, B and D presented against their actual values, when D is known. Solid blue and red lines correspond to the median of the true values under superoscillatory (blue squares) and plane wave illumination (red circles), while the red and blue colored bands indicate the corresponding interquartile (IQR) ranges (see also Supplementary Information). In the case of unknown position, panels (d-f) show the retrieved values of A, B and D presented against their actual values. Retrieved values for size C are similar to size A.

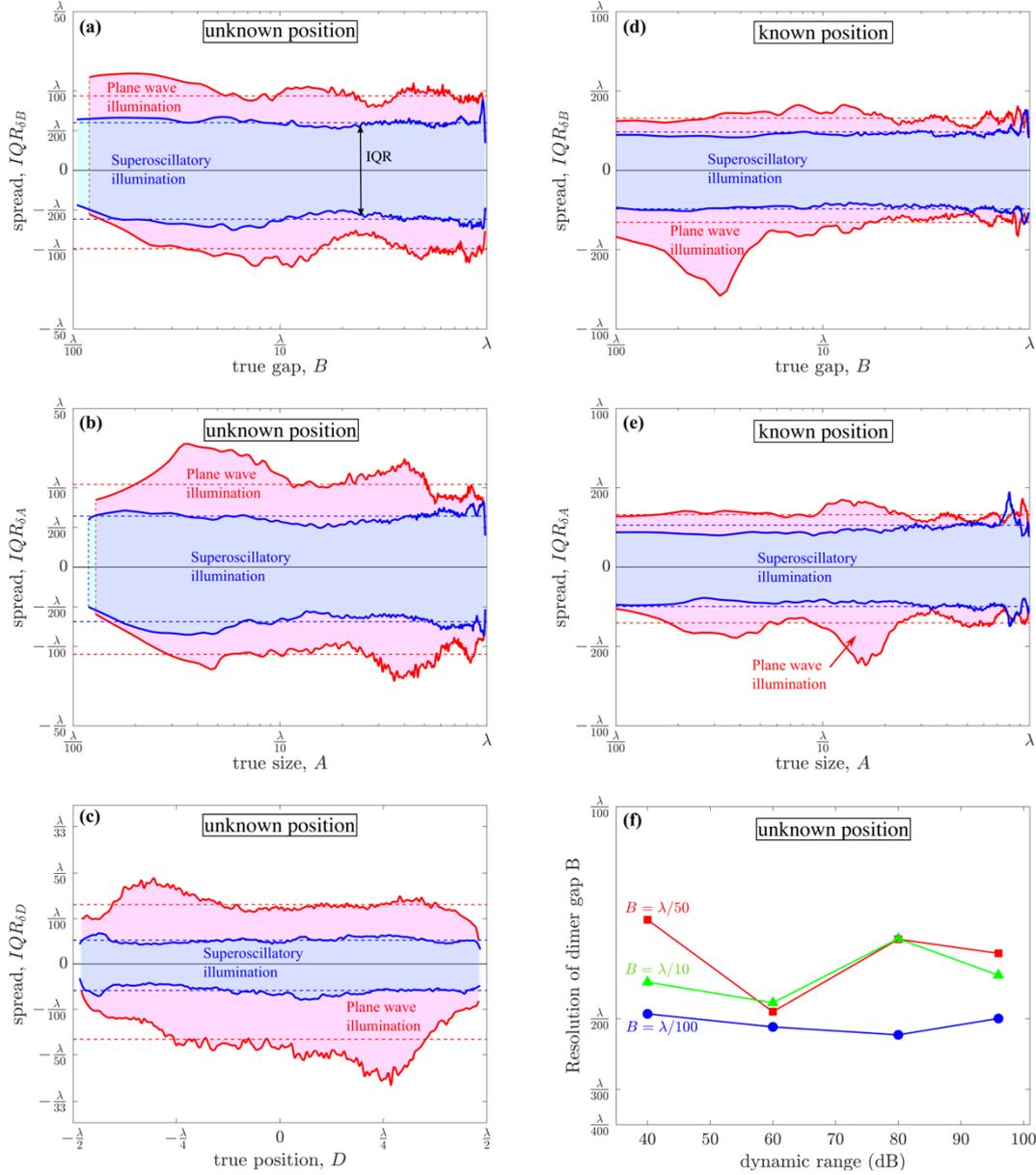

**Fig 3. Resolution of the Deeply Subwavelength Superoscillatory Imaging.** IQRs of measured values of the dimer parameters, gap, $B$ (a,d), element size, $A$ (b,e), and position, $D$ (c), during numerical imaging experiments with unknown (a-c) and known (d-e) dimer position. Red and blue colored regions correspond to plane wave and superoscillatory illumination, respectively, while red and blue solid lines mark the first and the third quartiles of the corresponding error distributions. The horizontal dotted lines indicate the average value of the IQRs over the range of the true values of the respective parameter. The vertical dotted lines in panels (a) and (b) indicate the parameter true value below which the network returns predominantly negative, non-physical values. (f) Dependence of resolution (in dimer gap $B$) as a function of the dynamic range of the photodetector.

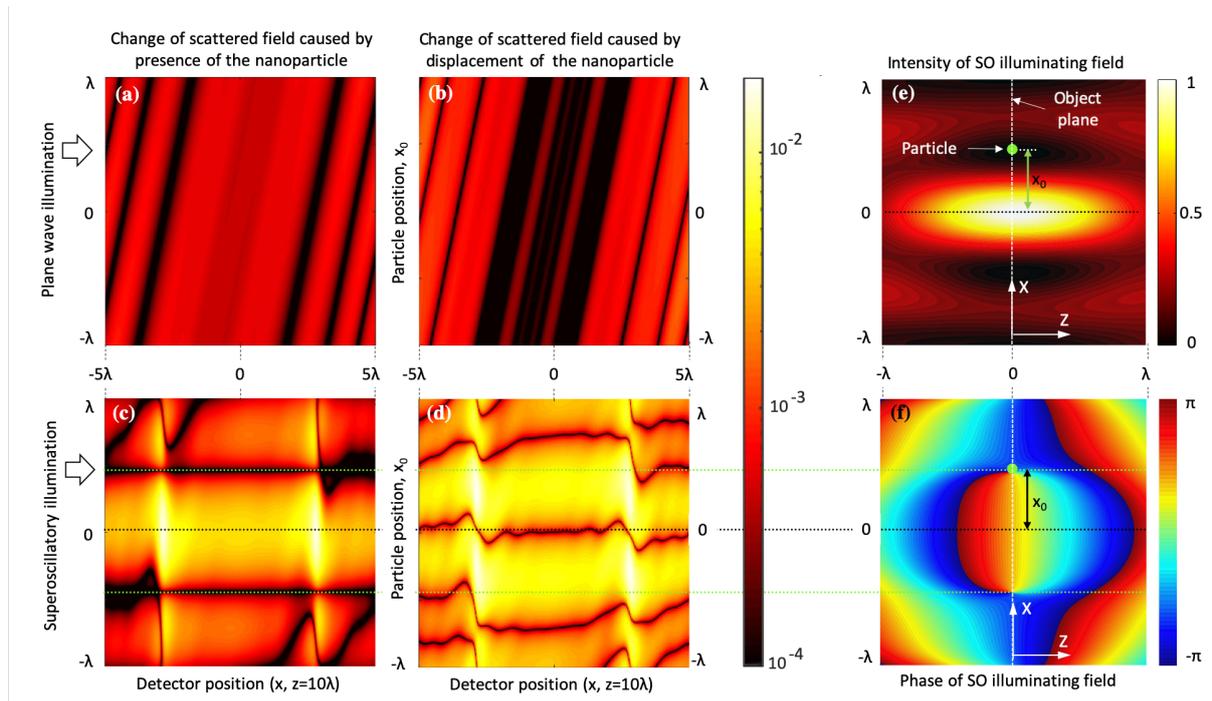

**Fig. 4. Sensitivity of far-field intensity patterns on presence and position of absorbing nanoparticle.** Plates (a) and (c) show normalized change of the scattered field intensity profile caused by presence of the nanoparticle. Plates (b) and (d) show normalized change of the scattered field intensity profile caused by shift of the nanoparticle on $\lambda/2000$ along *x* direction. Plates (a) and (b) correspond to a plane wave illumination; plates (c) and (d) illustrate illumination with superoscillatory field. Maps (e) and (f) show intensity and phase profiles of the illuminating superoscillatory field, where light propagates along the positive z-axis.

Supplementary Information for

# Unlabelled Far-field Deeply Subwavelength Superoscillatory Imaging (DSSI)

T. Pu, V. Savinov, G. Yuan, N. Papasimakis, N. I. Zheludev

Correspondence to: zheludev@soton.ac.uk

## Error evaluation

We quantify the error in the estimation of each of the four geometric parameters of the dimer *(A,B,C,D)* based on a large number (770,000) of scattering events. In each scattering event, the dimer parameters are selected in the range $\lambda/500<A,B,C<\lambda$ and $-\lambda/2<D<\lambda/2$ according to a random uniform distribution. We consider two closely related but distinct approaches of defining errors of the retrieval process, see Fig. S1. In both cases, a retrieval event is represented by the true gap value of the measured parameter, e.g. *B*, and its retrieved value $B+\delta B$.

In the first approach, we examine the spread of measured values of the dimer parameter for a given true value of the parameter**.** Here we define bins for true values and examine the distribution of retrieved values, $B+\delta B$, and corresponding errors, $\delta B$, within each bin. An example of the distribution is presented in Fig. S1a where the true gap value *B* is around $\lambda/50$. The distribution of the errors for the retrieval attempts can be characterized by the median value $\widetilde{\delta B}^R$ and the corresponding interquartile range $IQR^R_{\delta B}$, defined as the range between the second and third quartiles of the distribution containing 50% of all attempts. This procedure is repeated for all bins of true gap size, *B*. A similar procedure is followed for the dimer parameters *A, C* and *D*.

In the second approach, we examine the spread of true values of the dimer parameter for a given retrieved value of this parameter**.** Here we define the bins along the retrieved values and examine the distribution of true values, *B,* and corresponding errors, $\delta B$, within each bin. The distribution is characterized by the median $\widetilde{\delta B}^T$ and $IQR^T_{\delta B}$, see Fig. S1b. The corresponding resolution is defined as $IQR^T_{\delta B}/2$. Again, a similar procedure is followed for the dimer parameters *A, C* and *D*.

Medians and IQRs calculated according to the first approach are presented in Fig. S2, while results presented in the main text are calculated according to the second approach, see Figs. 2-3 in the main text. From these graphs one can see that both approaches return similar systematic offsets and resolution powers.

**Deeply Subwavelength Superoscillatory Imaging through noise**

Here we examine sensitivity of the DSSI method to noise. In a practical implementation of the method, noise can arise either as detection noise or due to unwanted scattering and interference effects. We model both of these effects by introducing an effective "noise field", $E_n$, at the detector plane, which takes values according to a zero mean Gaussian distribution with standard deviation σ. Thus, in the scenario of imaging a dimer, the total electric field at the detector plane will be the sum of the field, $E_s$, scattered by the dimer and the effective noise field, $E_n$. The corresponding total light intensity will be: $I = |E_s + E_n|^2$, or equivalently $I = |E_s|^2 + |E_n|^2 + 2Re[E_s E_n^*]$. The first term in this equation is the intensity of the field scattered by the dimer, the second term represents the detector's noise, while the third term accounts for the interference effects between light scattered from the dimer and any unwanted scattering. Assuming $E_s \gg E_n$, we quantify the noise level by the ratio $\eta = \frac{\sigma}{\max(|E_s|)}$, where $\max(|E_s|)$ is the maximum value of the modulus of the electric field at the detector plane.

Figure S3 shows the effect of noise on the retrieval of the dimer geometrical parameters. The noise results in the increase of divergence (bias) of the median lines (Fig. S3a&b) from the line of perfect imaging (black line in Figs. S3). In the case of 5% noise, substantial divergence occurs for *A*<λ/77 and *B*<λ/65, while for noise levels of 10%, the median diverges for *A*<λ/45 and *B*<λ/60.

The effects of noise on resolving power is presented in Figs. S3d-f. Here, increasing the noise level leads to gradual decrease of resolution. However, for all measured parameters resolution at 5% and 10% noise level remains at deeply subwavelength level, i.e. better than ~λ/70 and ~λ/55 correspondingly. This illustrates a remarkable resilience of the deconvolution process considering that we account for the interference phase related effects in the noise without providing any phase information to the network.

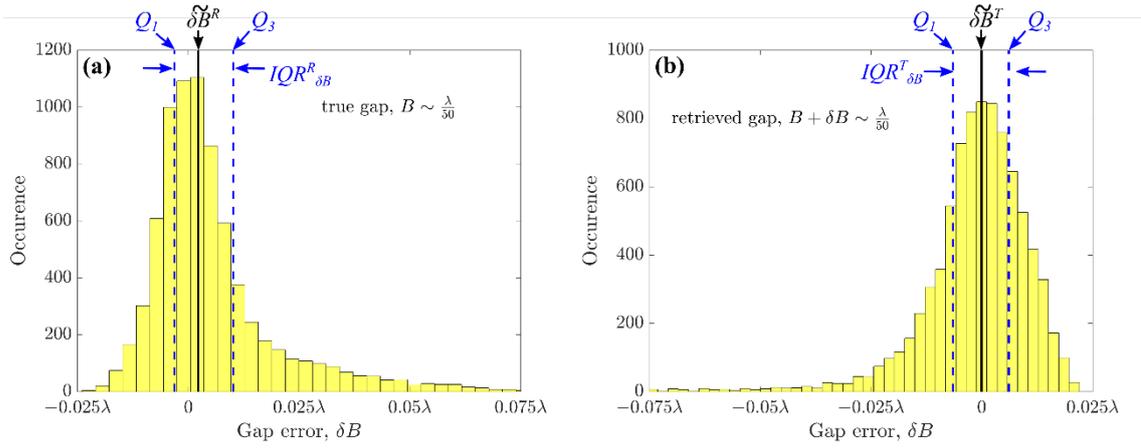

**Fig S1. Evaluation of imaging errors.** Panels (a) and (b) illustrate the two different ways in which errors of measuring the dimer gap $B$ can be defined, namely given a true parameter value (panel (a)) or a retrieved parameter value (panel (b)). The measurement events are placed in the bins of corresponding true value in panel (a) and corresponding retrieved value in panel (b). The panels show histograms of characteristic distributions of errors for $B \sim \lambda/50$. The solid black line marks the median ($\widetilde{\delta B}^R$,) of the error distribution. The interquartile ranges $IQR^R_{\delta B}$ and $IQR^T_{\delta B}$ are defined as the ranges between the corresponding first and third quartiles, which include 50% of the error values.

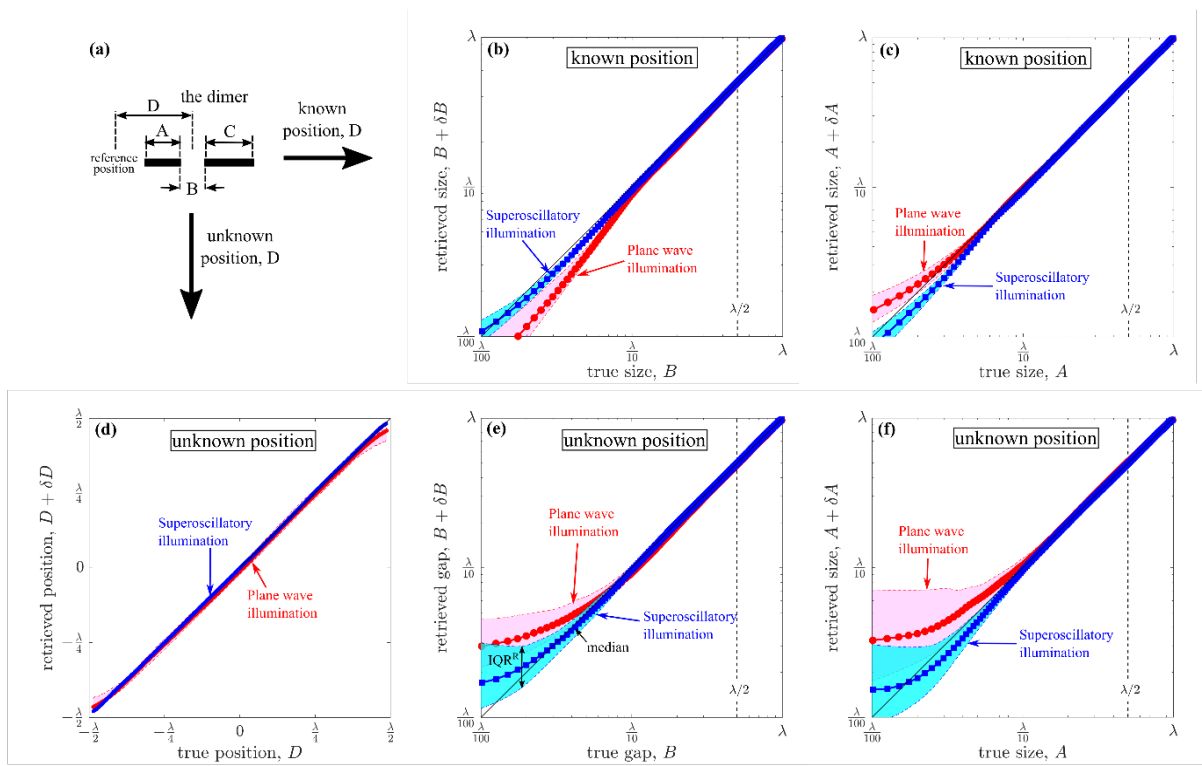

**Fig S2. Imaging errors for dimer imaging**. The retrieved values of the dimer parameter for a given true value of the parameters are presented. (a) Schematic of the dimer and its geometric parameters. Panels (b) and (c) show measurement results if location $D$ of the dimer is known. Solid blue and red lines correspond to the median of the retrieved values under superoscillatory (blue squares) and plane wave illumination (red circles), while the red and blue colored bands indicate the corresponding interquartile (IQR) ranges. In the case of unknown position of the dimer, panels (d) and (f) show the retrieved values of $A$, $B$ and $D$ against their actual values. Median values and IQRs are calculated according to the first approach as described in the Supplemenetary Information.

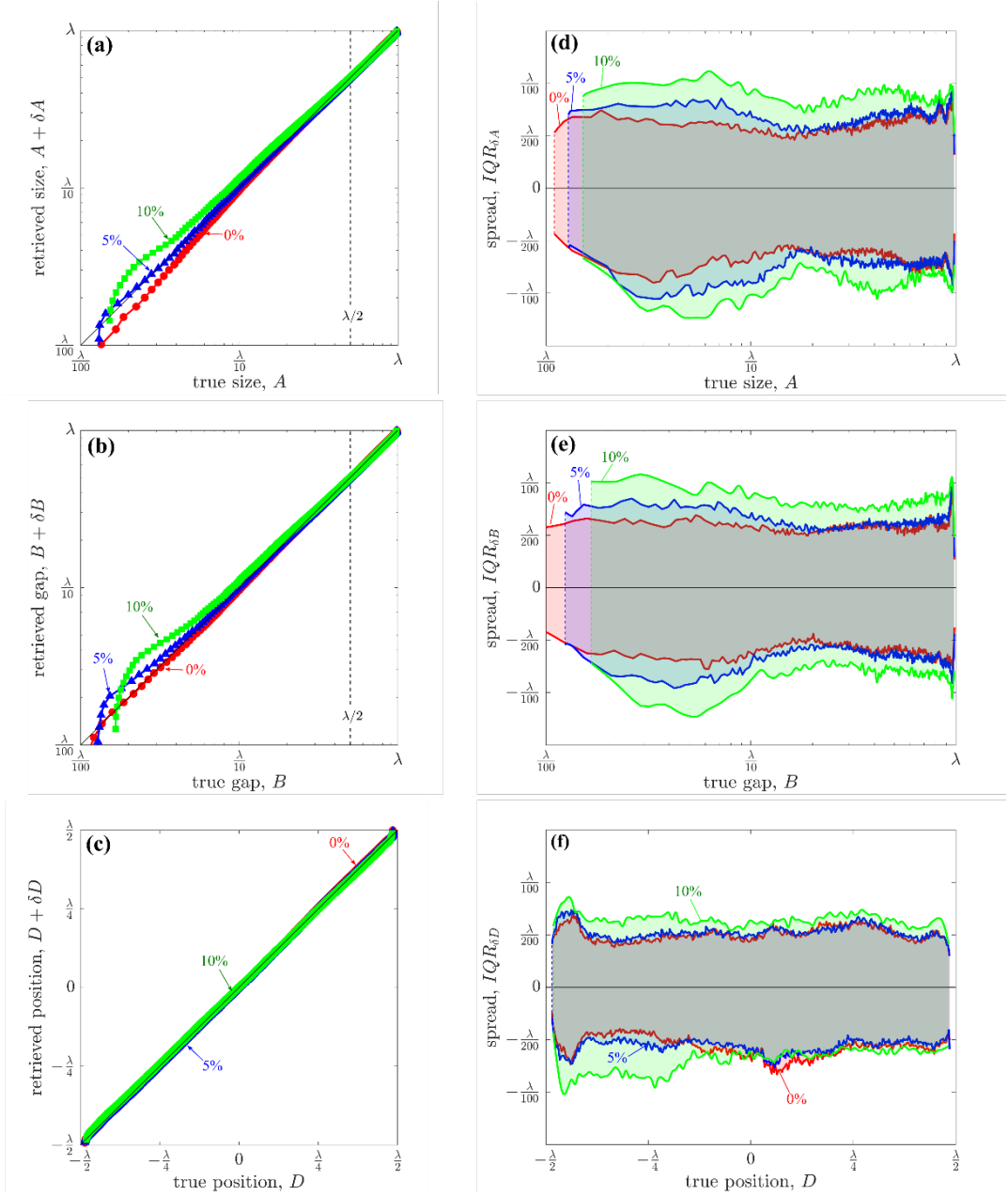

**Fig S3. Resilience of Deeply Subwavelength Superoscillatory Imaging to noise.** The figure shows the effects of different levels of noise $\eta$ on the retrieved dimer element size $A$ (a,b), gap $B$ (b,e), and position $D$ (c,f). Panels (a-c) show the median of retrieved values for $\eta = 0\%$ (red circles), 5% (blue triangles) and 10% (green squares). Red, blue, and green colored regions in panels (d), (e) and (f) show the spread (IQR) of errors for $\eta = 0\%$, 5%, and 10%, respectively. Median values and IQRs are calculated according to the second approach as described in Supplementary Information.

**Table S1. Resolution under superoscillatory and plane wave illumination.** The values in square brackets (for unknown position and under superoscillatory illumination) correspond to the resolution in the presence of 5% noise.

|  | Superoscillatory illumination | | Plane wave illumination | |
| --- | --- | --- | --- | --- |
|  | Unknown position | Known position | Unknown position | Known position |
| **Resolution in dimer element size, *A*** | 0.0133λ (λ/75) [0.0142 (λ/71)] | 0.0045λ (λ/222) | 0.0214λ (λ/47) | 0.0064λ (λ/156) |
| **Resolution in dimer gap, *B*** | 0.0122λ (λ/82) [0.0130 (λ/77)] | 0.0042λ (λ/238) | 0.0192λ (λ/52) | 0.0061λ (λ/164) |
| **Resolution in position, *D*** | 0.0111λ (λ/90) [0.0108 (λ/92)] |  | 0.0297λ (λ/34) |  |